\documentclass[aps,prl,twocolumn,superscriptaddress,showpacs]{revtex4}

\usepackage{graphicx}       
\usepackage{dcolumn}        
\usepackage{bm}             
\usepackage{amsmath}
\usepackage{amssymb}
\begin{document}


\title{Precise Measurement of the $\bm{\pi^+\to\pi^0 e^+\nu}$
Branching Ratio}

\newcommand*{\uva}{Department of Physics, University of Virginia,
                   Charlottesville, VA 22904-4714, USA}
\newcommand*{\psii}{Paul Scherrer Institut, Villigen PSI, CH-5232,
                    Switzerland}
\newcommand*{\dubna}{Joint Institute for Nuclear Research, RU-141980
                    Dubna, Russia}
\newcommand*{\swierk}{Institute for Nuclear Studies, PL-05-400 Swierk,
                    Poland}
\newcommand*{\tbilisi}{Institute for High Energy Physics, Tbilisi
                    State University, GUS-380086 Tbilisi, Georgia}
\newcommand*{\asu}{Department of Physics and Astronomy, Arizona State
                   University, Tempe, AZ 85287, USA}
\newcommand*{\irb}{Rudjer Bo\v{s}kovi\'c Institute, HR-10000 Zagreb,
                   Croatia} 
\affiliation{\uva}
\affiliation{\psii}
\affiliation{\dubna}
\affiliation{\swierk}
\affiliation{\tbilisi}
\affiliation{\asu}
\affiliation{\irb}

\author{D.~Po\v{c}ani\'c}\email[Corresponding author:] 
                   {pocanic@virginia.edu}\affiliation{\uva}
\author{E.~Frle\v{z}}\email[Corresponding author:] 
                   {frlez@virginia.edu}\affiliation{\uva}
\author{V.~A.~Baranov}\affiliation{\dubna}
\author{W.~Bertl}\affiliation{\psii}
\author{Ch.~Br\"onnimann}\affiliation{\psii}
\author{M.~Bychkov}\affiliation{\uva}
\author{J.~F.~Crawford}\affiliation{\psii}
\author{M.~Daum}\affiliation{\psii}
\author{N.~V.~Khomutov}\affiliation{\dubna}
\author{A.~S.~Korenchenko}\affiliation{\dubna} 
\author{S.~M.~Korenchenko}\affiliation{\dubna}
\author{T.~Kozlowski}\affiliation{\swierk}
\author{N.~P.~Kravchuk}\affiliation{\dubna}
\author{N.~A.~Kuchinsky}\affiliation{\dubna} 
\author{W.~Li}\affiliation{\uva}
\author{R.~C.~Minehart}\affiliation{\uva}
\author{D.~Mzhavia}\affiliation{\dubna} 
\author{B.~G.~Ritchie}\affiliation{\asu}
\author{S.~Ritt}\affiliation{\psii}
\author{A.~M.~Rozhdestvensky}\affiliation{\dubna} 
\author{V.~V.~Sidorkin}\affiliation{\dubna}
\author{L.~C.~Smith}\affiliation{\uva}
\author{I.~Supek}\affiliation{\irb} 
\author{Z.~Tsamalaidze}\affiliation{\tbilisi}
\author{B.~A.~VanDevender}\affiliation{\uva}
\author{Y.~Wang}\affiliation{\uva}
\author{H.-P.~Wirtz}\altaffiliation[Presently at: ]{Philips 
Semiconductors AG, CH-8045  Z\"urich, Switzerland}\affiliation{\psii}
\author{K.~O.~H.~Ziock}\affiliation{\uva}


\date{9 July 2004}

\begin{abstract}

Using a large acceptance calorimeter and a stopped pion beam we have
made a precise measurement of the rare $\pi^+\to \pi^0 e^+\nu
\,(\pi_\beta)$ decay branching ratio.  We have evaluated the branching
ratio by normalizing the number of observed $\pi_\beta$ decays to the
number of observed $\pi^ +\to e^+\nu$ ($\pi_{e2}$) decays.  We find
the value of $\Gamma (\pi^+\to \pi^0 e^+\nu) / \Gamma(\text{total}) =
[1.036\pm 0.004\,\rm(stat.)  \pm 0.004\,(syst.)  \pm
0.003\,(\pi_{e2})]\times 10^{-8}$, where the first uncertainty is
statistical, the second systematic, and the third is the $\pi_{e2}$
branching ratio uncertainty.  Our result agrees well with the Standard
Model prediction.

\end{abstract}

\pacs{13.20.Cz, 11.40.-q,14.40.Aq}

\maketitle

The rare pion beta decay, $\pi^+\to\pi^0e^+\nu$ (branching ratio
$R_{\pi\beta} \simeq 1 \times 10^{-8}$), is one of the most basic
semileptonic electroweak processes.  It is a pure vector transition
between two spin-zero members of an isospin triplet, and is therefore
analogous to superallowed Fermi (SF) transitions in nuclear beta
decay.  Due to its simplicity and accuracy, the theory of Fermi beta
decays is one of the most precise components of the Standard Model
(SM) of electroweak interactions.

The CVC hypothesis \cite{Ger55,Fey58} and quark-lepton universality
relate the rate of pure vector beta decay (both pion and nuclear) to
that of muon decay via the Cabibbo-Kobayashi-Maskawa (CKM) quark
mixing matrix element $V_{ud}$ \cite{Cab63,Kob73}.  Including loop
corrections, $\delta$, the rate of pion beta decay is given by
\cite{Kal64,Sir78}:
\begin{equation}
  \Gamma_{\pi\beta} = 
    \frac{G^2_\mu|V_{ud}|^2}{30\pi^3}
              \left( 1 - \frac{\Delta}{2M_+} \right)^3
    \Delta^5f(\epsilon,\Delta)(1+\delta) \,,   \label{eq:pb_rate_tree}
\end{equation}
where $G_\mu$ is the Fermi weak coupling constant, $\Delta= M_+ -
M_0$, $\epsilon = (m_e/\Delta)^2$, $M_+$, $M_0$, and $m_e$ are the
masses of the $\pi^+$, $\pi^0$, and the electron, respectively, while
$f$, the Fermi function, is given by
\begin{align}
  f(\epsilon,\Delta) &= 
     \sqrt{1-\epsilon} \Bigg[1 - \frac{9}{2}\epsilon - 4\epsilon^2
                              \label{eq:fermi_fn}   \\ \nonumber
    & +\frac{15}{2}\epsilon^2 \ln\left( \frac{1+\sqrt{1-\epsilon}}
        {\sqrt{\epsilon}} \right)
        - \frac{3}{7}\frac{\Delta^2}{(M_+ + M_0)^2} \Bigg]\ . 
\end{align}
The main experimental source of uncertainty in $\Gamma_{\pi\beta}$
amounts to just 0.05\,\%; it comes from the measurement of $\Delta$
\cite{Cra91}.  The combined radiative and short-range physics
corrections amount to $\delta \simeq 0.033$ and are exceptionally well
controlled, yielding an overall theoretical uncertainty of
$\Gamma_{\pi\beta}$ of $\lesssim 0.1\,$\%
\cite{Sir78,Sir82,Jau01,Cir02}.  Hence, pion beta decay presents an
excellent means for a precise experimental determination of the CKM
matrix element $V_{ud}$, hindered only by the low branching ratio of
the decay.

The CKM quark mixing matrix has a special significance in modern
physics as a cornerstone of a unified description of the weak
interactions of mesons, baryons and nuclei.  In a universe with three
quark generations the $3\times 3$ CKM matrix must be unitary, barring
certain classes of hitherto undiscovered processes not contained in
the Standard Model.  Thus, an accurate experimental evaluation of the
CKM matrix unitarity provides an independent check of possible
deviations from the SM.  As the best studied element of the CKM
matrix, $V_{ud}$ plays an important role in all tests of its
unitarity.  However, evaluations of $V_{ud}$ from neutron decay have,
for the most part, not been consistent with results from nuclear SF
decays \cite{PDG04}.  Clearly, a precise evaluation of $V_{ud}$ from
pion beta decay, the theoretically cleanest choice, is of interest.

The most precise measurement of the pion beta decay rate on record was
made by McFarlane et al., at LAMPF by detecting in-flight $\pi^+$
decays in the LAMPF $\pi^0$ spectrometer \cite{McF85}.  This work
reported $\Gamma_{\pi\beta} = 0.394 \pm 0.015\,$s$^{-1}$, which is an
order of magnitude less precise than the theoretical description of
the same process.  Hence, we initiated the PIBETA experiment, a
program of precise measurements of the rare pion and muon decays at
rest, chief among them being the pion beta decay, at the Paul Scherrer
Institute (PSI), Switzerland \cite{Poc92}.

\begin{figure}[tb]
\hbox to\hsize{\hss
\includegraphics[width=\hsize]{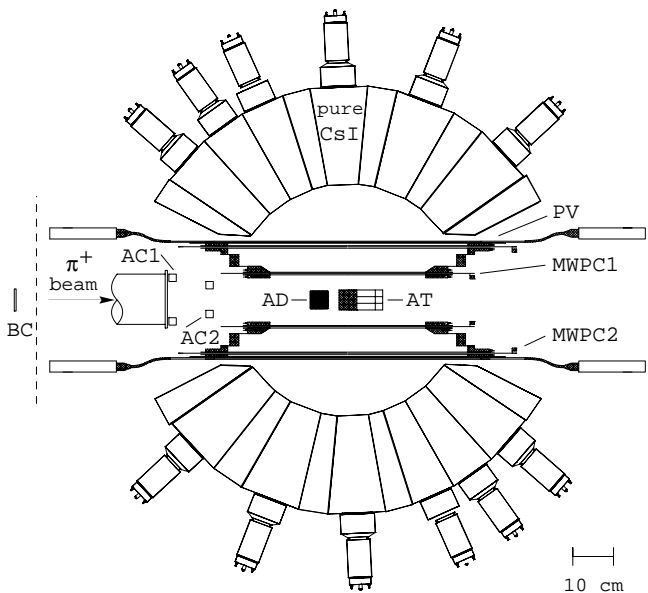}  
\hss}
\caption{A schematic cross section of the PIBETA detector system.
Symbols denote: BC--thin upstream beam counter, AC1,2--active beam
collimators, AD--active degrader, AT--active target, MWPC1,2--thin
cylindrical wire chambers, PV--thin 20-segment plastic scintillator
barrel.  BC, AC1, AC2, AD and AT detectors are also made of plastic
scintillator.}
\label{fig:xsect}
\end{figure}

In this Letter we present an analysis of the $\pi^+ \to \pi^0 e^+\nu$
decay events recorded with the PIBETA apparatus from 1999 to 2001.  We
tuned the $\pi$E1 beam line at PSI to deliver $\sim 10^6$ $\pi^+$/s
with $p_\pi \simeq 113\,$MeV/c.  The pions were slowed in an active
degrader detector (AD) and stopped in a segmented 9-element active
target (AT), both made of plastic scintillator.  The major detector
systems are shown in a schematic drawing in Fig.~\ref{fig:xsect}.
Energetic charged decay products are tracked in a pair of thin
concentric multiwire proportional chambers (MWPCs) and a thin
20-segment plastic scintillator barrel veto detector (PV).  Both
neutral and charged particles deposit most (or all) of their energy in
a spherical electromagnetic shower calorimeter consisting of 240
elements made of pure CsI.  The entire detector system, its response
to positrons, photons and protons, energy and time resolution, signal
definitions, along with other relevant details of our experimental
method, are described at length in Ref.~\cite{Frl03a}.

The measurement relies on detecting the $\pi^0 \to \gamma\gamma$ decay
which immediately follows a pion beta decay event.  The two photons
are emitted nearly back-to-back, with about 67\,MeV each.  Thus, the
experiment is set to record all large-energy (above the $\mu \to
e\nu\bar{\nu}$ endpoint) electromagnetic shower pairs occurring in
opposing detector hemispheres during a $\sim$\,180\,ns long ``pion
gate'', $\pi$G (non-prompt two-arm events).  The $\pi$G is timed so as
to include a sample of pile-up events preceding the pion stop.  In
addition, we record a large prescaled sample of non-prompt single
shower (one-arm) events.  Using these minimum-bias sets, we extract
the $\pi_\beta$ and $\pi_{e2}$ event sets, the latter for branching
ratio normalization.  In a stopped pion experiment these two channels
have nearly the same detector acceptance, and have much of the
systematics in common.

A full complement of twelve fast analog triggers comprising all
relevant logic combinations of one- or two-arm, low- or high
calorimeter threshold (labeled {\sc ht} and {\sc lt}, respectively),
prompt and delayed (with respect to $\pi^+$ stop time), as well as a
random and a three-arm trigger, were implemented in order to obtain
maximally comprehensive and unbiased data samples.

\begin{figure}[b]
\hbox to\hsize{\hss\includegraphics[width=\hsize]{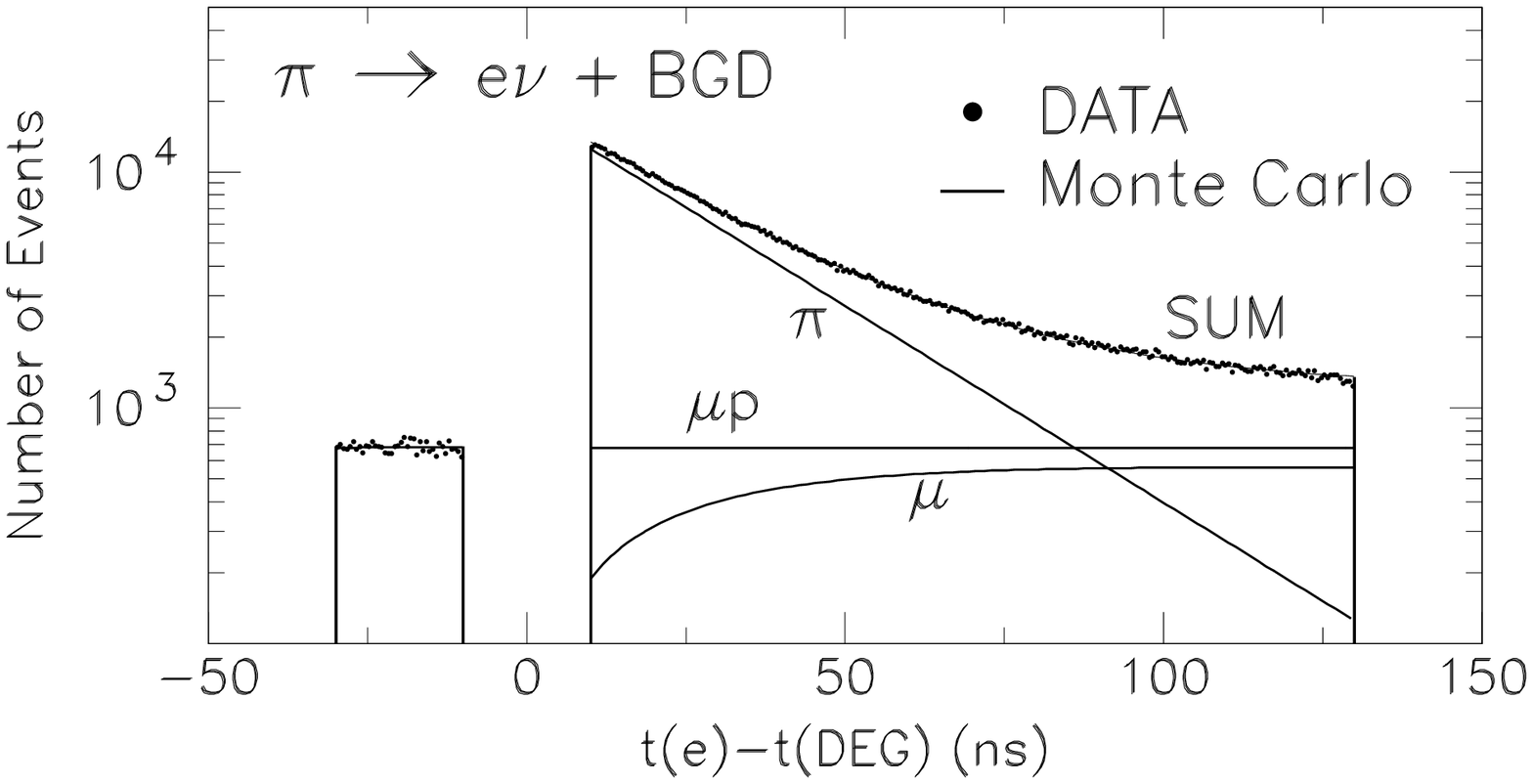}\hss} 
\hbox to\hsize{\hss\includegraphics[width=\hsize]{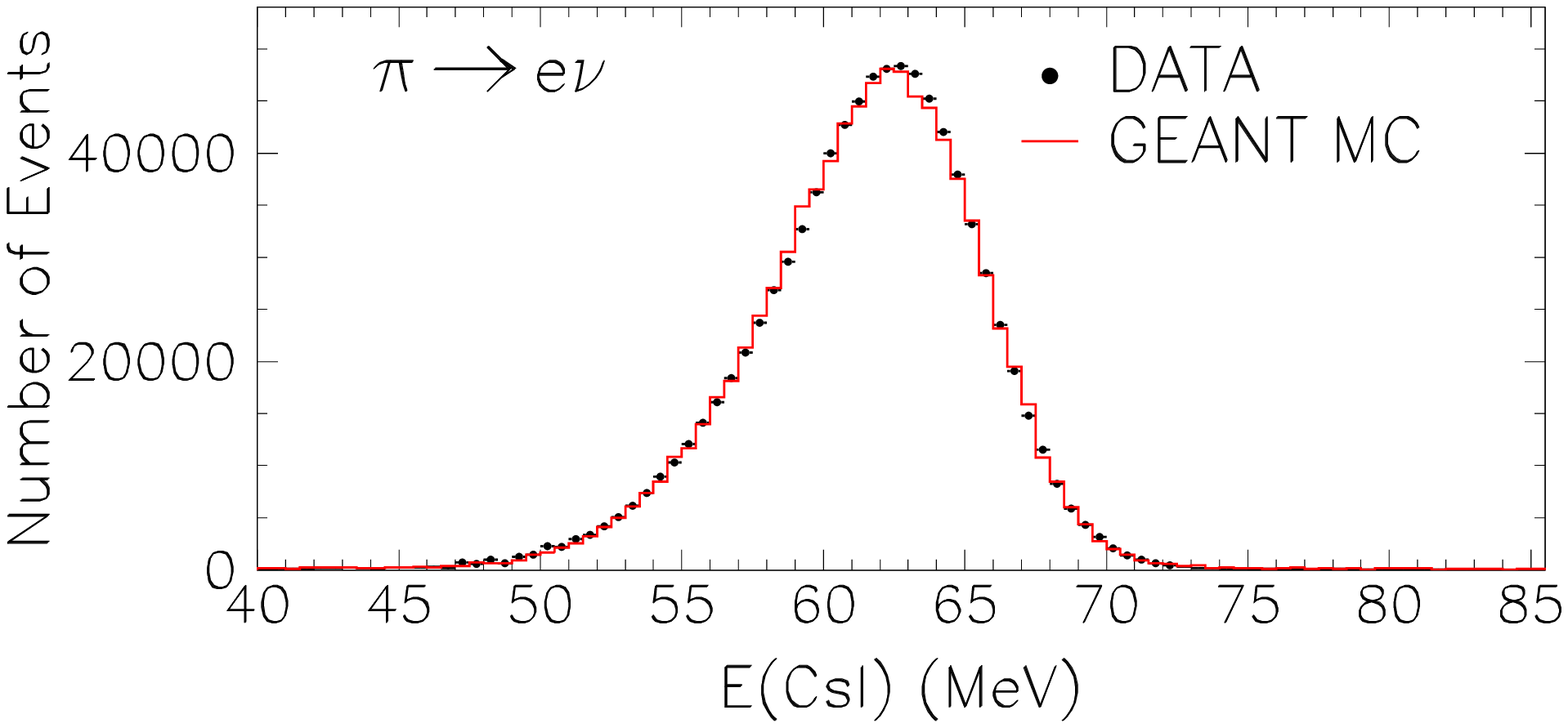}\hss} 
\caption{Top panel: A typical histogram of time differences between
the beam pion stop, $t$({\sc deg}), and 1-arm {\sc ht} event time,
$t$(e), (dots), compared with a sum of the Monte Carlo-simulated
responses for $\pi_{e2}$ decay ($\pi$), muon decay ($\mu$), and muon
pile-up events ($\mu$p).  The $\pi_{e2}$ pile-up background, being
much lower, is off scale in the plot.  Prompt events are suppressed.
Bottom panel: CsI calorimeter energy spectrum for the $\pi_{e2}$ decay
events, after background subtraction.  }
\label{fig:pe2}
\end{figure}

Signal definition and accurate counting of the $\pi_{e2}$ events for
normalization present a major challenge in this work.  As in all
previous studies, our $\pi_{e2}$ data include undiscriminated
soft-photon $\pi_{e2\gamma}$ events.  Due to positron energy
straggling in the target, accidental coincidences of multiple muon
decay events, and the calorimeter energy resolution function, the
$\pi_{e2}$ events are superimposed on a non-negligible muon decay
background.  This background was removed by fitting the measured $e^+$
timing spectra with the functions for pion decay (signal), muon decay
(background), plus the associated pile-ups (see Fig.~\ref{fig:pe2}
top).  We also extracted the absolute $\pi_{e2}$ branching ratio using
this method and normalizing to the number of pion stops in the target.
The results were in agreement with the recommended Particle Data Group
(PDG) value \cite{PDG04} at a sub-percent level, with the uncertainty
dominated by the systematics of the stopped pion counting.  The latter
is absent in our determination of $R_{\pi\beta}$.  The $\pi_{e2}$
energy spectrum after background subtraction is given in
Fig.~\ref{fig:pe2} bottom.  The statistical uncertainty of the extracted
number of $\pi_{e2}$ events, $N_{\pi e2}$, is negligible.

\begin{figure}[t]
\hbox to\hsize{\hss\includegraphics[width=\hsize]{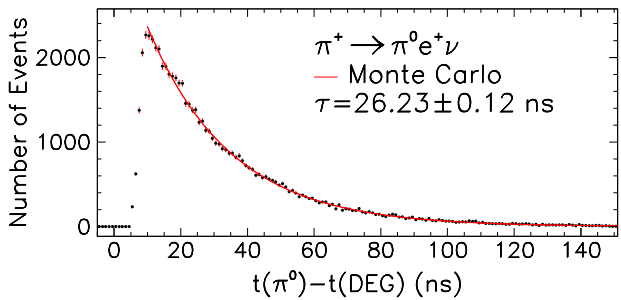} 
               \hss}
\caption{Histogram of time differences between the beam pion stop and
the $\pi_\beta$ decay events (dots); curve: pion lifetime.  }
\label{fig:pb_tim}
\end{figure}

\begin{figure}[b]
\hbox to\hsize{\hss\includegraphics[width=\hsize]{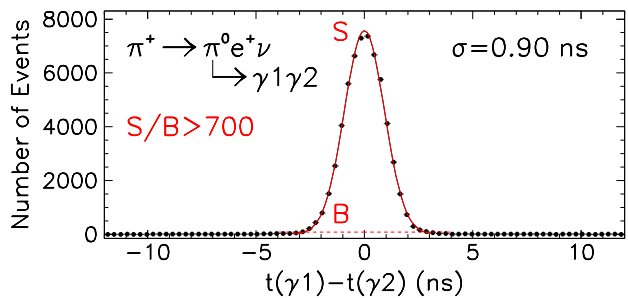} 
               \hss}
\caption{Histogram of $\gamma$-$\gamma$ time differences for
$\pi_\beta$ decay events (dots); curve: fit with a Gaussian function
plus a constant. }
\label{fig:pb_sn}
\end{figure}

\begin{figure}[t]
\hbox to\hsize{\hss\includegraphics[width=\hsize]{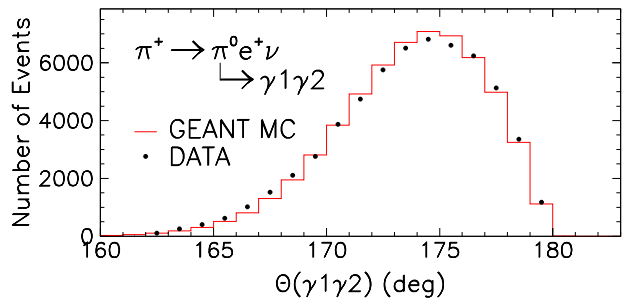} 
               \hss}
\caption{Histogram of the $\gamma$-$\gamma$ opening angle in $\pi_\beta$
decay.  }
\label{fig:pb_th12}
\end{figure}

The $\pi_\beta$ signal definition was more straightforward, as seen in
Figs.~\ref{fig:pb_tim} and \ref{fig:pb_sn}, which show the pion decay
time spectrum and $\gamma$-$\gamma$ relative timing histogram,
respectively, for $\pi_\beta$ events, both free of backgrounds.
Finally, the histogram of recorded $\gamma$-$\gamma$ opening angles
for pion beta events, shown in Fig.~\ref{fig:pb_th12}, provides a
sensitive test of the systematics related to the geometry of the beam
pion stopping distribution, an important contributor to the acceptance
uncertainty.

The $\pi_\beta$ branching ratio $R_{\pi\beta}$ was evaluated from
\begin{equation}
  {R}_{\pi\beta} 
   = \frac{N_{\pi\beta}}{N_{\pi^+}\, f_{\rm \pi G}\,
      A_{\pi\beta}^{\rm HT}\, \tau_{\rm l}\, f_{\rm CPP} \,
      f_{\rm D} \, f_{\rm ph} } \, ,   \label{eq:R_pibeta}
\end{equation}
where $N_{\pi\beta}$ is the number of detected $\pi_\beta$ events,
$N_{\pi^+}$ is the number of the decaying $\pi^+$'s, $f_{\rm\pi G}$ is
the delayed pion gate fraction, $A_{\pi\beta}^{\rm HT}$ is the {\sc ht}
detector acceptance evaluated by {\tt GEANT} simulation, $\tau_{\rm
l}$ is the detector live time, $f_{\rm CPP}$ is the correction due to
the charged particle (CP) veto system pile-up, $f_{\rm D}$ is the
$\pi^0$ Dalitz decay correction, and $f_{\rm ph}$ is the photonuclear
absorption correction.

The $\pi\to e\nu$ branching ratio $R_{\pi e2}$
is given by
\begin{equation}
  R_{\pi e2}
    = \frac{N_{\pi e2}\, p_{\pi e2}}
           { N_{\pi^+}\, f_{\rm \pi G}
            \, A_{\pi e2}^{\rm HT}\,
            \tau_{\rm l}\, \epsilon_{\rm PV}\, 
            \epsilon_{\rm C1}\, \epsilon_{\rm C2}}\, ,
                               \label{eq:R_pie2}
\end{equation}
where $p_{\pi e2}$ is the prescaling factor applied to $\pi_{e2}$
triggers, $A_{\pi e2}^{\rm HT}$ is the high-threshold detector
acceptance for $\pi\to e\nu$ decay events, including radiative
corrections, while $\epsilon_{\rm PV}$, $\epsilon_{\rm C1}$, and
$\epsilon_{\rm C2}$ denote the plastic veto and wire chamber
efficiencies, respectively.  Clearly, taking the ratio
$R_{\pi\beta}/R_{\pi e\nu}$ leads to cancellations of many
common factors, apart from small corrections taking into account
slight differences in thresholds, trigger timing (two-arm vs.\
one-arm), weighting of the efficiencies, and similar effects.  Most
importantly, $N_{\pi^+}$, the number of stopped pions drops out.  The
main sources of uncertainty are listed with their values in
Table~\ref{tab:pb_unc}.

\begin{table}[b]

\caption{Summary of the main sources of uncertainty in the extraction
of $R_{\pi\beta}$, given in \% (see text for discussion).  }
\label{tab:pb_unc}
\begin{ruledtabular}
\begin{tabular}{llcc}
  Type & Quantity  & Value & Unc.\ (\%)  \\
\hline \\[-2ex]
 external:
    & $\Gamma(\pi_{e2})$       & $1.230 \times 10^{-4}$ & 0.33 \\ 
    & $\Gamma(\pi^0\to\gamma\gamma)$ 
                               &  0.9880                & 0.03 \\
    & $\pi^+$ lifetime         &  26.033\,s             & 0.02 \\[1ex]
 \multicolumn{3}{l}{combined external uncertainties:}   & 0.33 \\[1ex]
 internal: 
    & $N_{\pi e2}$ systematic  &  $6.779 \times 10^8$  & 0.19  \\
    & $A_{\pi\beta}^{\rm HT}/A_{\pi e2}^{\rm HT}$  
                               &  0.9432               & 0.12  \\
    & $r_{\rm\pi G}=f_{\rm\pi G}^{\pi\beta}/f_{\rm\pi G}^{\pi e2}$ 
                               &  1.130                & 0.26  \\
    & $\pi_\beta$ accid.\ bgd.  &  0.00               & $<0.1$  \\
    & $f_{\rm CPP}$ correction &  0.9951               & 0.10  \\
    & $f_{\rm ph}$ correction  &  0.9980               & 0.10  \\[1ex]
 \multicolumn{3}{l}{combined internal uncertainties:}  & 0.38  \\[1ex]
 statistical: & $N_{\pi\beta}$ &  64,047               & 0.395 \\
\end{tabular}
\end{ruledtabular}
\end{table}

As the external systematic uncertainties are self-ex\-pla\-na\-to\-ry,
we turn to the internal ones.  The systematic uncertainty in $N_{\pi
e2}$ comes from the muon-decay background subtraction discussed above,
and reflects the propagated error limits of the method.  The precision
of $A_{\pi\beta}^{\rm HT}/A_{\pi e2}^{\rm HT}$ is dominated by the
uncertainty of the $x$-$y$-$z$ distribution of pion stops in the
target.  The latter was determined with better than 50\,$\mu$m
accuracy by tomographic back-tracing of $\pi_{e2}$ and muon decay
positrons into the target \cite{WLi04}.  Corrections due to the
undetected low portions of the $e$ and $\gamma$ energy spectra in the
calorimeter (the energy ``tail'') contribute weakly to the acceptance
uncertainty due to strong correlations between the energy responses to
the two decay channels.  This experiment has a unique advantage over
its predecessors: it measures branching ratios as well as differential
angular and energy distributions of decay products for all rare pion
and muon decays simultaneously.  This provides multiple redundant
consistency checks of the evaluated and simulated acceptances (cf.,
e.g., Ref.~\cite{Frl03b}).  In the present analysis the largest
internal contribution to the systematic uncertainty comes from the
ratio of gate fractions, $r_{\rm\pi G}$, due to our decision to
include even the earliest $\pi_\beta$ decay events, thus maximizing the
useful event statistics.  An alternative method involves a sharp cut
at, say, 10\,ns after the pion stop time, which reduces the systematic
error, along with the $\pi_\beta$ event statistics, yielding, however,
a consistent $R_{\pi\beta}$.  The inherent resolution in the zero time
point is excellent---it relies on the prompt A$(\pi^+,\pi^0)$B signal
and the accelerator rf pulse, providing timing calibration at the
level of $\sim$\,20\,ps or better, and room for further improvement of
the $r_{\rm\pi G}$ precision.  The pile-up correction $f_{\rm CPP}$
was evaluated using a random trigger, and confirmed by simulations.
We modified our GEANT3 code to calculate the photonuclear correction
$f_{\rm ph}$, and conservatively assigned it a 50\,\% uncertainty
(details are given in Ref.~\cite{pb_web}).  Efficiencies
$\epsilon_{\rm PV}$, $\epsilon_{\rm C1}$, and $\epsilon_{\rm C2}$, not
listed in Table \ref{tab:pb_unc}, were measured with an accuracy of
0.01\,\% \cite{WLi04}.

Using the above method and the PDG 2004 recommended value of $R_{\pi
e2}^{\text{exp}}=1.230(4)\times 10^{-4}$~\cite{PDG04}, we extract the
pion beta decay branching ratio, our main result:
\begin{equation}
   R^{\text{exp}}_{\pi\beta} = \rm 
      [1.036 \pm 0.004\,(stat) \pm 0.005\,(syst)] \times 10^{-8}\,,
   \label{eq:pb_br}
\end{equation}
or, in terms of the decay rate,
\begin{equation}
   \Gamma^{\text{exp}}_{\pi\beta} = \rm
       [0.3980 \pm 0.0015(stat) \pm 0.0019(syst)]\,s^{-1}\,,
\end{equation}
which represents a six-fold improvement in accuracy over the
previous measurement~\cite{McF85}.  Alternatively, the
normalization can be tied to the theoretical value $R_{\pi
e2}^{\text{the}} = (1.2352 \pm 0.0005)\times 10^{-4}$~\cite{Mar93}
which would increase the extracted $R^{\text{exp}}_{\pi\beta}$ by
$0.42\,$\% to $1.040 \times 10^{-8}$.  In a direct evaluation of the
pion beta decay branching ratio using Eq~(\ref{eq:R_pibeta}), i.e.,
normalizing to the number of beam pion stops, we obtain
$R_{\pi\beta}\cdot 10^8 = 1.042 \pm 0.004(\text{stat}) \pm
0.010(\text{syst})$, consistent with our main result given in
Eq.~(\ref{eq:pb_br}).

Whether scaled to the experimental or theoretical $R_{\pi e2}$, our
result for $R^{\text{exp}}_{\pi\beta}$ is in excellent agreement with
predictions of the SM and CVC given the PDG recommended value range
for $V_{ud}$ \cite{PDG04}:
\begin{equation}
    R^{\text{SM}}_{\pi\beta} 
          = (1.038 - 1.041) \times 10^{-8} \quad \text{(90\% C.L.)}\,,
                                    \label{eq:pb_sm_pred}
\end{equation}
and represents the most accurate test of CVC and Cabibbo universality
in a meson to date.  Our result confirms the validity of the radiative
corrections for the process at the level of $4\sigma_{\text{exp}}$,
since, excluding loop corrections, the SM would predict
$R^{\text{no\ rad.\ corr.}}_{\pi\beta} = (1.005 - 1.007) \times
10^{-8}$ at 90\,\% C.L. 

Using our result, Eq.~(\ref{eq:pb_br}), we can calculate a new value
of $V_{ud}$ from pion beta decay, $V_{ud}^{\text{(PIBETA)}} =
0.9728(30)$, which is in excellent agreement with the PDG 2004
average, $V_{ud}^{\text{(PDG'04)}} = 0.9738(5)$.  We will continue to
improve the overall accuracy of the $\pi_\beta$ decay branching ratio
to $\sim$\,0.5\,\% by further refining the experiment simulation and
analysis, and by adding new data.

\begin{acknowledgments}

We thank W.~A.~Stephens, Z.~Hochman, and the PSI experimental support
group for invaluable help in preparing and running the experiment.
This work has been supported by the U.S. National Science Foundation,
the U.S. Department of Energy, the Paul Scherrer Institute, and the
Russian Foundation for Basic Research.

\end{acknowledgments}

\bibliography{pb_prl1b}

\end{document}